\documentclass[reprint,amsmath,amssymb,aps]{revtex4-1}

\usepackage{graphicx}% Include figure files
\usepackage{dcolumn}% Align table columns on decimal point
\usepackage{bm}% bold math
\usepackage{color}

\begin{document}

\preprint{APS/123-QED}

\title{MoTe$_{2}$ as a natural hyperbolic material across the visible and the ultraviolet region}

\author{Saeideh Edalati-Boostan}
 \email{edalati@physik.hu-berlin.de}
\affiliation{Physics Department and IRIS Adlershof, Humboldt-Universit\"at zu Berlin, 12489 Berlin, Germany
}%
\author{Caterina Cocchi}%
 \email{caterina.cocchi@physik.hu-berlin.de}
\affiliation{Physics Department and IRIS Adlershof, Humboldt-Universit\"at zu Berlin, 12489 Berlin, Germany
}%
\author{Claudia Draxl}%
 \email{claudia.draxl@physik.hu-berlin.de}
\affiliation{Physics Department and IRIS Adlershof, Humboldt-Universit\"at zu Berlin, 12489 Berlin, Germany
}%

\date{\today}% It is always \today, today,

\begin{abstract}
Hyperbolic materials are of particular interest for the next generation of photonic and opto-electronic devices. Since artificial metamaterials are intrinsically limited by the size of their nanostructured components, there has been a {\it hunt} for natural hyperbolic materials in the last few years. In a first-principles work based on density-functional theory and many-body perturbation theory, we investigate the fundamental dielectric response of MoTe$_{2}$ in monolayer, bilayer, and bulk form, and find that it is a natural type-II hyperbolic material with low losses between 3 and 6 eV. Going from the monolayer to the bulk, the energy window of hyperbolic dispersion is blue-shifted by a few tenths of an eV. We show that excitonic effects and optical anisotropy play a major role in the hyperbolic behavior of MoTe$_{2}$. Our results confirm the potential of layered materials as hyperbolic media for opto-electronics, photonics, and nano-imaging applications.
\end{abstract}

%\keywords{TMDC, hyperbolic dispersion, optical properties}

\maketitle

\section{\label{sec:level1}Introduction}
Hyperbolic metamaterials (HMM) are multifunctional systems for waveguiding, imaging, sensing, as well as quantum and thermal engineering, which operate beyond the limits of conventional devices \cite{smith2004negative, belov2005canalization, poddubny2013hyperbolic, podolskiy2005strongly, noginov2013focus, balmain2002resonance}. Their name comes from the \textit{hyperboloid} shape of the isofrequency surface formed by the dispersion relation of transversally polarized electromagnetic waves

 \begin{equation}
  \frac{k_{\parallel}^{2}}{\epsilon_{\perp}(\omega)}+\frac{k_{\perp}^{2}}{\epsilon_{\parallel}(\omega)}= \frac{\omega^{2}}{c^{2}}, 
\label{eqn:dispersion}
\end{equation}  
where $k$, $\omega$, and $c$ are the wave vector, the frequency of radiation, and the velocity of light in vacuum, respectively. 

The in-plane and out-of-plane components of the dielectric tensor with respect to the anisotropy axis $z$ are indicated by $\epsilon_{\parallel}\equiv\epsilon_{xx} = \epsilon_{yy}$ and $\epsilon_{\perp}\equiv\epsilon_{zz}$, respectively. The hyperbolicity condition is fulfilled  by these components having opposite sign, i.e. $\epsilon_{\parallel}\epsilon_{\perp}<0$ \cite{poddubny2013hyperbolic, caldwell2014jd, guo2012applications, dai2014s}. Depending on the sign of the individual  components, hyperbolic media are classified as two types, namely type-I characterized by $\epsilon_{\perp} < 0, \epsilon_{\parallel} > 0$, and type-II by $\epsilon_{\parallel} < 0, \epsilon_{\perp} > 0$ ~\cite{korobkin2010measurements}. HMM are typically formed by artificial nanostructures, including alternating layers of dielectric and metallic media ~\cite{poddubny2013hyperbolic}. However, their often complicated fabrication process, the relatively large size of the metallic components, and the electron scattering due to the internal interfaces limit their performance \cite{Tong2018, BOLTASSEVA20081}. The hunt for natural hyperbolic materials has therefore captured increasing attention in the last few years~\cite{korzeb2015compendium,gomez2016flatland}. Unfortunately, the strict structural requirements that enable hyperbolic dispersion from the ultraviolet to the terahertz region have been shown to be fulfilled only by a limited number of materials, including graphite~\cite{sun2011indefinite,iorsh2013hyperbolic} and a few selected cuprates~\cite{sun2014indefinite}, ruthenates~\cite{korzeb2015compendium}, mineral tetradymites \cite{esslinger2014tetradymites,shekhar2019fast}, oxides~\cite{eaton2018vo,zheng2019mid}, and electrides~\cite{guan2017tunable}.
  
Hyperbolic dispersion with negative permittivity is generally obtained by plasmon polaritons  or phonon polaritons  \cite{brar2014hybrid}. In addition, a third type of polaritons, namely exciton polaritons (EPs), are characterized by low energy loss and an  operating frequency window in the visible region. Hyperbolic behavior in the presence of EPs, arising from bound electron-hole pairs and large optical anisotropy, promises new perspectives for opto-electronic applications with visible light. For example, it has recently been reported  \cite{guo2018hyperbolic} in two-dimensional (2D) perovskites, caused by the presence of anisotropic excitons.

Strong excitonic effects and large optical anisotropy are exhibited concurrently by a few classes of materials. Transition metal dichalcogenides (TMDCs) are one of them and, as such, are considered promising candidates as natural hyperbolic materials across the visible region~\cite{gjerding2017band,gjerding2017layered}. Due to their layered structure, they are naturally prone to optical anisotropy \cite{gong2017electronic,wang+17prl,PhysRevB.88.245309}, while their low dimensionality gives rise to localized states and, hence, strong excitonic effects \cite{berkelbach2013theory, chernikov2014exciton, li2014measurement, wang2018colloquium, druppel2018electronic} that dominate their optical spectra \cite{qiu+13prl,steinhoff2014influence, moody2015intrinsic, thygesen2017calculating,lau+19prm}. Hyperbolic behavior in these systems at UV frequencies has been predicted by calculations based on the independent-particle approximation \cite{gjerding2017layered}. 

Considering that transition-metal tellurides, such as MoTe$_{2}$, have a smaller band gap compared to their disulfide and diselenide counterparts, one could expect a red-shift of the hyperbolicity window to the visible spectrum. Moreover, the fact that MoTe$_{2}$ shows only a moderate decrease in the photoluminescence intensity by increasing  number of layers \cite{lezama2015} may grant hyperbolic dispersion even in multilayers and bulk crystals, contrary, for example, to perovskites, which are reported to be hyperbolic only in the monolayer and bilayer limit \cite{guo2018hyperbolic}. This fascinating hypothesis requires theoretical validation in a framework that properly accounts for electron-hole correlation effects. In a first-principles study based on many-body perturbation  theory, we demonstrate that MoTe$_{2}$ is a natural type-II hyperbolic material with low losses in a window of a few eV from the upper boundary of the visible to the UV region. Strong excitonic effects and large optical anisotropy result in hyperbolic dispersion regardless of the number of layers.

\section{\label{sec:level1}Theoretical background}

This work is carried out in the framework of density functional theory (DFT)~\cite{hohe-kohn64pr,kohn-sham65pr} and many-body perturbation theory~\cite{onid+02rmp}. DFT calculations are performed to obtain the equilibrium lattice constants and to minimize the interatomic forces in the unit cell, with interlayer van der Waals interactions described by the DFT-D2 scheme~\cite{grimme2006semiempirical}. The Kohn-Sham (KS) eigenstates and eigenenergies are used as a starting point for the subsequent steps. 

We adopt the $GW$ approach in the single-shot $G_0W_0$ approximation~\cite{hedi65pr, hybertsen1985first} to obtain the quasi-particle (QP) energies as:
\begin{equation}
 \varepsilon_{n\bf k}^{QP}= \varepsilon_{n\bf k}^{KS}+Z_{n\bf k}[\Re \Sigma_{n\bf k} (\varepsilon_{n\bf k}^{KS})-V_{n\bf k}^{xc}], \label{eqn:QP}
\end{equation}                                
where $ \varepsilon_{n\bf k}^{KS}$, $\Sigma_{n\bf k}$, $V_{n\bf k}^{xc}$, and  $Z_{n\bf k}$ are the KS eigenvalues, the electronic self-energy, the exchange-correlation potential, and the renormalization factor, respectively. The dielectric tensors are computed from the solution of the Bethe-Salpeter equation (BSE)~\cite{strinati1988application}, the equation of motion of the electron-hole correlation function~\cite{strinati1988application}. In the eigenvalue form, the BSE reads~\cite{rohl-loui00prb,pusc-ambr02prb}: 
\begin{equation}
\sum_{v'c'\bf k'}H_{vc{\bf k},v'c'\bf k'}^{BSE} A_{v'c'\bf k'}^{\lambda}= E^{\lambda} A_{vc\bf k}^{\lambda},
  \label{eqn:BSE}
\end{equation}    
where, in the Tamm-Dancoff approximation the effective two-particle Hamiltonian for a spin-degenerate system is defined as
\begin{equation}
H^{BSE} = H^{diag} + 2 H^{x} + H^{d}.
\label{eqn:BSE_H}
\end{equation}
This expression includes the diagonal term, $H^{diag}$, that accounts for the contributions of transitions energies between non-interacting QP states. The exchange term, $H^{x}$, describes the short-range repulsive exchange interaction between electron and hole while the direct term, $H^{d}$, considers the electron-hole attraction driven by the statically screened Coulomb potential. The solution of the BSE provides all the ingredients to calculate the macroscopic dielectric function, $\epsilon_{M}(\omega)$, whose imaginary part
\begin{equation}
  \Im \epsilon_{M} = \frac{8\pi^{2}}{\Omega}\sum_{\lambda} |\mathbf{t}^{\lambda}|^{2} \delta(\omega-E^{\lambda}). 
  \label{eqn:Im}
\end{equation}    
describes the optical absorption of the material. In Eq.~\ref{eqn:Im}, $t^{\lambda}$ are the transition coefficients of the individual excitations, obtained as:
\begin{equation}
\mathbf{t}^{\lambda} = \sum_{vc\bf k}A_{vc\bf k}^{\lambda}\frac{\langle v\bf k|\widehat{\bf p}|c \bf k \rangle}{\epsilon_{c\bf k}^{QP}-\epsilon_{v\bf k}^{QP}},
  \label{eqn:Tr}
\end{equation}  
including the BSE eigenvectors $A_{vc\bf k}^{\lambda}$ from Eq.~\ref{eqn:BSE}. The numerator of Eq.~\ref{eqn:Tr} contains the matrix elements of the momentum operator. The loss function in the bulk is given by $\Im \epsilon^{-1}_{M}$. 

In the bilayer (BL) and monolayer (ML) phases, the systems are simulated in supercells with a sufficiently large amount of vacuum in the out-of-plane direction to prevent unphysical interactions between neighboring replica. In order to exclude spurious contributions arising from the vacuum layer, we rescale $\epsilon_{M}$ by replacing $\Omega$, the volume of the entire supercell, in Eq.~\eqref{eqn:Im} with an effective volume, $\Omega_{eff}$, obtained as ~\cite{kathrin2014}
\begin{equation}
\Omega_{eff} = \frac{a_{eff}}{a_{z}}\Omega, 
  \label{eqn:volunme}
\end{equation}  
where $a_{eff}$ is the material thickness (i.e., 3.65 \AA{} in the ML and 10.65 \AA{} in the BL) while $a_{z}$=$a_{eff}$+$a_{vac}$ includes the thickness of the vacuum layer $a_{vac}$, equal to 18 \AA{}.

\section{\label{sec:level2}Computational details}

All calculations are carried out using \texttt{exciting}~\cite{gulans2014exciting}, a full-potential all-electron package for DFT and MBPT, implementing linearized augmented planewave (LAPW) methods. The sampling of the Brillouin zone is performed with Monkhorst-Pack \cite{monkhorst1976special} \textbf{k}-point meshes, with 18$\times$18$\times$1 (18$\times$18$\times$2) grids for ML and BL (bulk).  A cutoff of 3.3 bohr$^{-1}$ is used for  the planewave part of the basis. Values of 2.4 bohr and 2.5 bohr are taken for the muffin-tin radii of Mo and Te, respectively. Spin-orbit coupling (SOC) is included through the second-variational scheme~\cite{gulans2014exciting}. The generalized gradient approximation in the Perdew-Burke-Ernzerhof (PBE) parameterization~\cite{perdew1996generalized} is adopted for the exchange-correlation functional. All atoms are relaxed during the geometry optimization until the forces acting on any atom are smaller than 0.01 eV \AA$^{-1}$.  

In the $G_{0}W_{0}$ calculations~\cite{nabok2016accurate}, 100 empty bands and a \textbf{k}-mesh of 12$\times$12$\times$1 are used for ML and BL to compute the screened Coulomb interaction within the random-phase approximation (RPA). For the bulk, a 6$\times$6$\times$2 \textbf{k}-mesh and 500 empty bands are used. We did not employ a cutoff for the Coulomb potential, assuming the partial compensation between corrections of both QP and exciton binding energies~\cite{PhysRevB.99.161201}. 

BSE calculations ~\cite{vorwerk2019bethe} are performed within the Tamm-Dancoff approximation on top of the electronic structure that includes the QP correction for the fundamental gap as a scissors operator. In the ML and BL (bulk) the Brillouin zone is sampled by a 16$\times$16$\times$1 (16$\times$16$\times$2) \textbf{k}-mesh. 12 valence bands and 12 conduction bands are included in the transition space to obtain excitation energies and oscillator strengths for the ML by including SOC. In the BL and the bulk, 24 valence and 24 conduction bands are employed. These parameters provide excitation energies converged within 50 meV.

Input and output data are stored in the NOMAD Repository ~\cite{Draxl_2019} and are freely available for download at the following link: http://dx.doi.org/10.17172/NOMAD/2019.07.31-1.

\section{\label{sec:level1}Results}
\begin{figure}[th]
\centering
\includegraphics[width=0.9\linewidth]{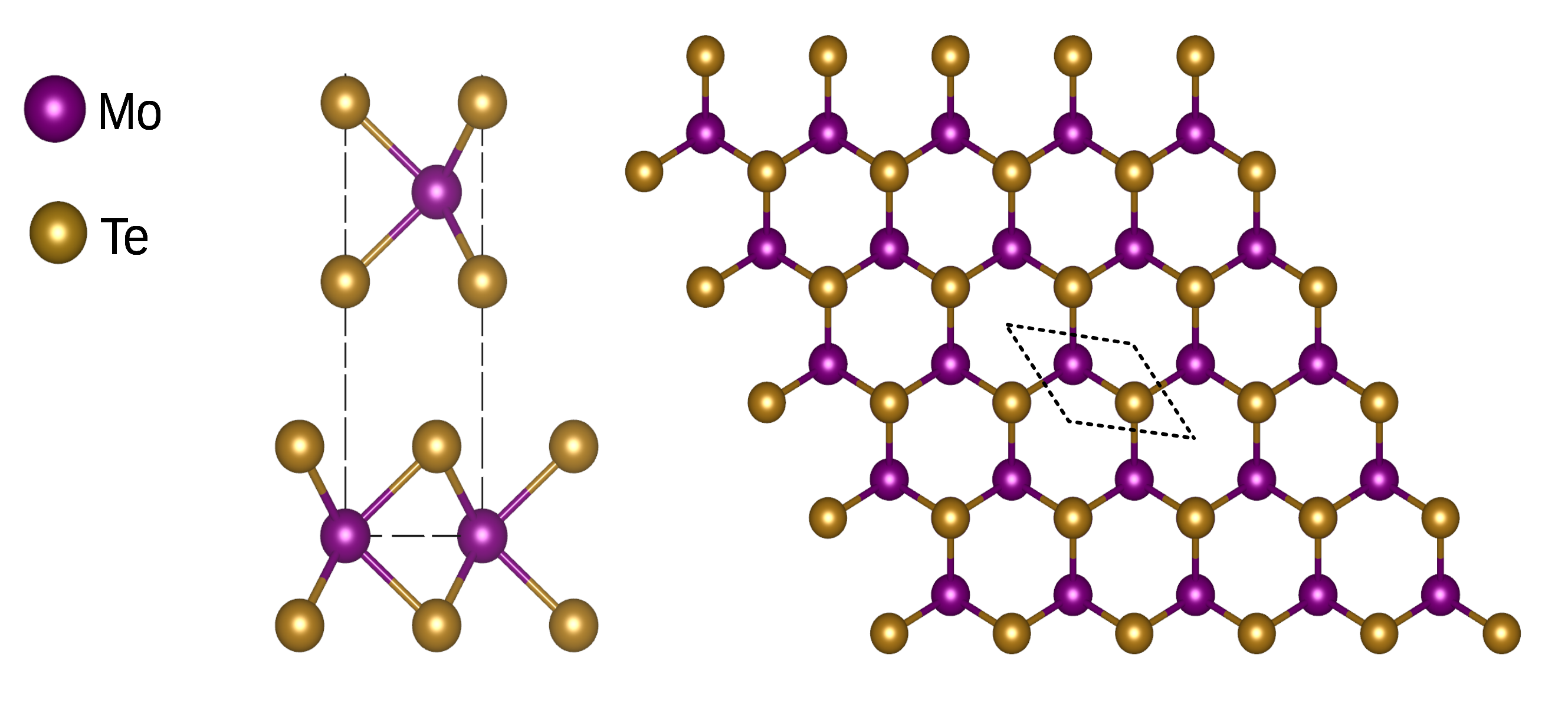}
  \caption{Side (left) and top (right) view of bulk 2H-MoTe$_{2}$. Mo atoms are indicated in purple, Te atoms in gold. The unit cell is marked by dashed lines. \label{fig:structure}}
\end{figure}
2H-MoTe$_{2}$ with AA' stacking geometry (point group D$_{3d}$) has been reported to be the most stable structure in the BL and bulk phases~\cite{he2014stacking}. In this eclipsed stacking, the Mo atom is located directly on top of the Te atoms of the next layer, as shown in Fig. \ref{fig:structure}. 

\begin{figure}[th]
\includegraphics[width=0.32\textwidth]{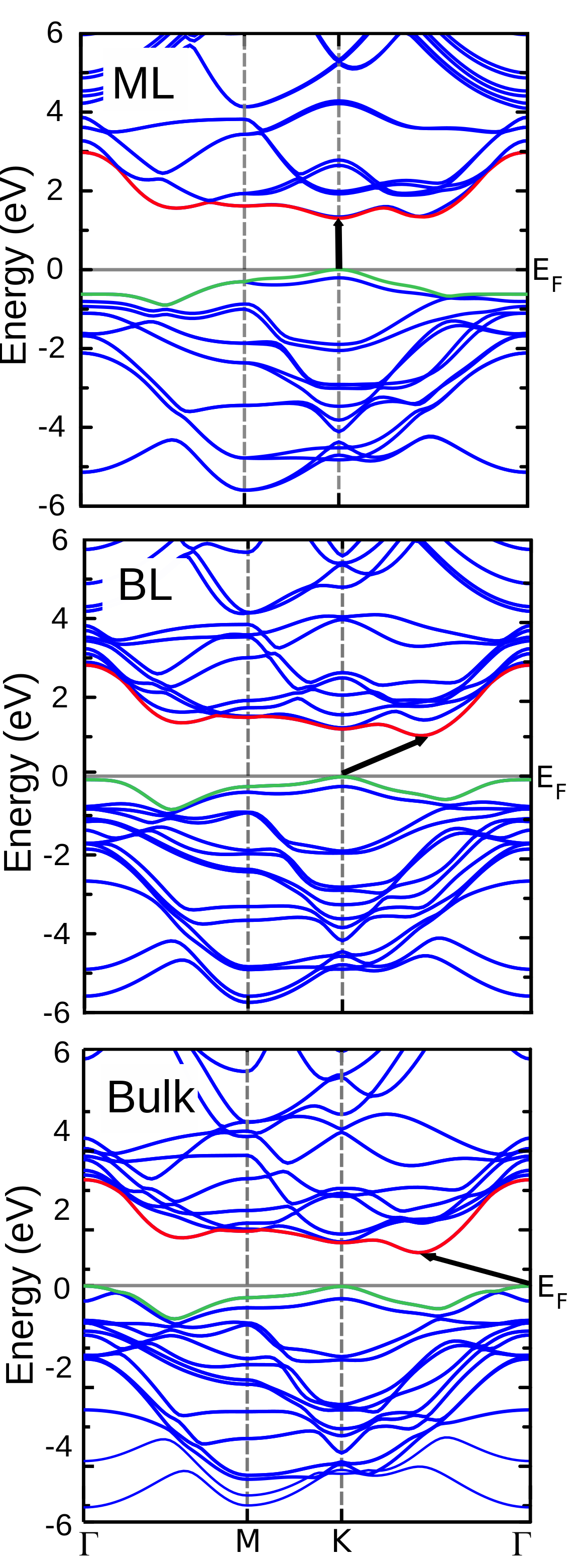}
\caption{Quasi-particle band structures of ML (top), BL (middle), and bulk (bottom) MoTe$_{2}$ including SOC. The highest occupied and lowest unoccupied bands are marked in green and red, respectively. The black arrows indicate the fundamental gap between the top of the valence band (set at zero energy) and the bottom of the conduction band. 
\label{fig:band}}
\end{figure}
In Fig.~\ref{fig:band} we show the QP band structures of MoTe$_{2}$ including SOC. Similar to its disulfide and diselenide counterparts \cite{gusakova2017electronic}, MoTe$_{2}$ has a direct band gap only in the ML limit, with the valance band maximum (VBM) and the conduction band minimum (CBm) both located at the high-symmetry point K. By increasing the number of layers, the VBM  shifts from K to $\Gamma$, and the CBm moves along the K-$\Gamma$ path, leading to an indirect band gap. The SOC causes a splitting of the valance bands at the high-symmetry point K by 205 meV, 239 meV, and 385 meV in ML, BL, and bulk, respectively. Our PBE results (not shown) are in overall agreement with  corresponding values reported in the literature \cite{bhattacharyya2012semiconductor,rasm-thyg15jpcc,Haastrup_2018}, yielding band gaps of 1.04 eV, 0.90 eV, and 0.72 eV in ML, BL, and bulk, respectively. The optical gaps, similar to the fundamental band gaps, decrease with increasing number of layers, ranging from 1.30 eV in the ML (equal to the fundamental gap) to 1.26 eV in the BL, and 1.15 eV in the bulk.  The smaller differences between the direct band gap in the ML (E$_{gap}^{QP}$ =  1.30 eV) and the indirect band  gaps in the BL and bulk (1.06 eV and 0.86 eV, respectively), compared to other TMDCs \cite{lezama2015}, explains the only moderate decrease of PL intensity upon increasing number of layers~\cite{froehlicher2016direct}. Hence, in this work we focus on 2H-MoTe$_{2}$.

\begin{figure*}[th]
\includegraphics[width=0.8\textwidth]{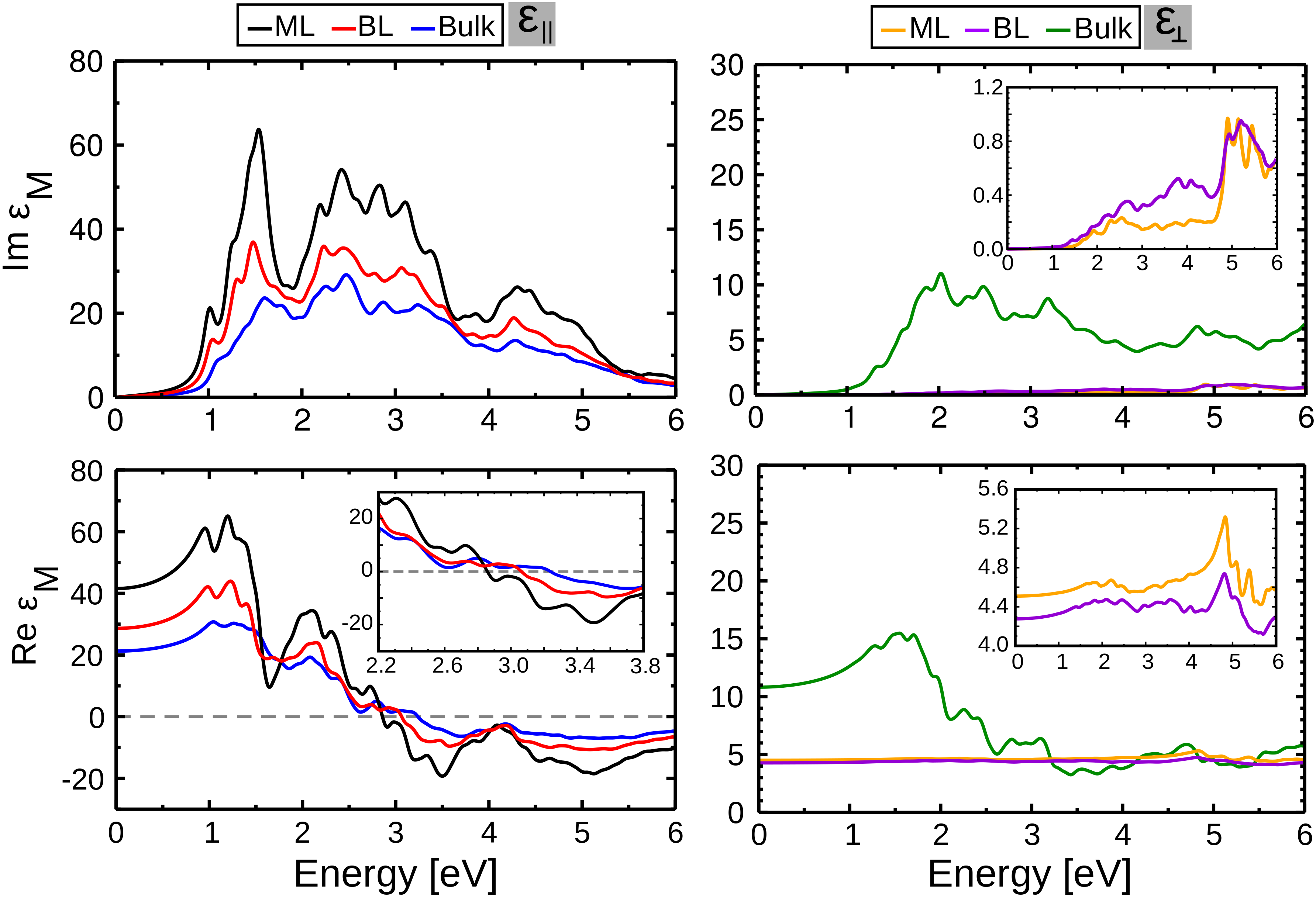}
\caption{ Imaginary (top) and real (bottom) parts of the in-plane (left) and the out-of-plane components (right) of the imaginary (top) and real part (bottom) of the macroscopic dielectric function of ML, BL, and bulk MoTe$_{2}$. The insets in the right panels magnify the regions where Im$\epsilon_M$ in the ML and BL is significantly smaller compared to the bulk. The inset in the lower left panel highlights the region in which Re$\epsilon_M$ changes sign. A Lorentzian broadening of 82 meV is applied to mimic excitation lifetimes.
\label{fig:Im_Re}}
\end{figure*}

The structural anisotropy of MoTe$_{2}$ is reflected in its dielectric response. In Fig. \ref{fig:Im_Re} we display the in-plane (left) and out-of-plane components (right) of the imaginary (top) and real part (bottom) of the macroscopic dielectric function of the ML, BL, and bulk phases. We start by inspecting the imaginary part, which provides information on the frequency-dependent absorption of the materials. The in-plane components of $\Im \epsilon_{M}$ in the ML and BL are intrinsically larger compared to the bulk. From the top left panel of Fig.~\ref{fig:Im_Re} we notice that the absorption onset, given by the transition between the highest occupied band and the lowest-unoccupied band at K, is at about 1 eV, regardless of the number of layers. This lowest-energy excitation is characterized by an exciton binding energy which decreases from about 300 meV in the ML, to 230 meV in the BL, to 70 meV in the bulk. This behavior is consistent with the screening enhanced by the increasing number of layers. The values are in good overall agreement with previous results obtained at the same level of theory~\cite{robert2016excitonic, Arora2017, Haastrup_2018}. Differences of the order of 100 meV in the exciton binding energies can be ascribed to the sensitivity of these calculations to the \textbf{k}-mesh and the number of states included in the transition space \footnote{As an example, reducing the \textbf{k}-grid from 16$\times$16$\times$1 to 12$\times$12$\times$1 in the bulk, considering 24 bands, the binding energy is increased from 69 meV to 71 meV. However, for 12 bands this value increases significantly to 480 meV and 620 meV using 16$\times$16$\times$1 and 12$\times$12$\times$1 \textbf{k}-grids, respectively.}.

The first exciton is localized within a single layer in all the structures, as found in TMDCs~\cite{qiu+13prl,Arora2017} as well as in other multilayer materials~\cite{aggo+18prb,pale+18excitons} and heterostructures~\cite{aggo+17jpcl,lau+19prm}. The in-plane component of $\Im \epsilon_M$ exhibits analogous features above the onset for all structures, with the oscillator strength decreasing systematically with the number of layers (see Fig.~\ref{fig:Im_Re}, top left  panel). A maximum centered at 1.5 eV and composed by several excitations is found in all the spectra, mainly originating from transitions between the highest occupied band and the lowest unoccupied band along the path K-M, close to K. The next maximum is at about 2.5 eV and is formed by transitions between the second highest occupied band and the lowest-unoccupied band and also from the highest occupied band and the second lowest unoccupied band around M. Additional peaks in the three spectra appear at the high-energy edge of the visible region at about 3 eV and stem from transitions between the highest occupied band and lowest unoccupied band around $\Gamma$. In the UV region, the in-plane component decreases monotonically. 

\begin{figure*}[th]
\includegraphics[width=0.8\textwidth]{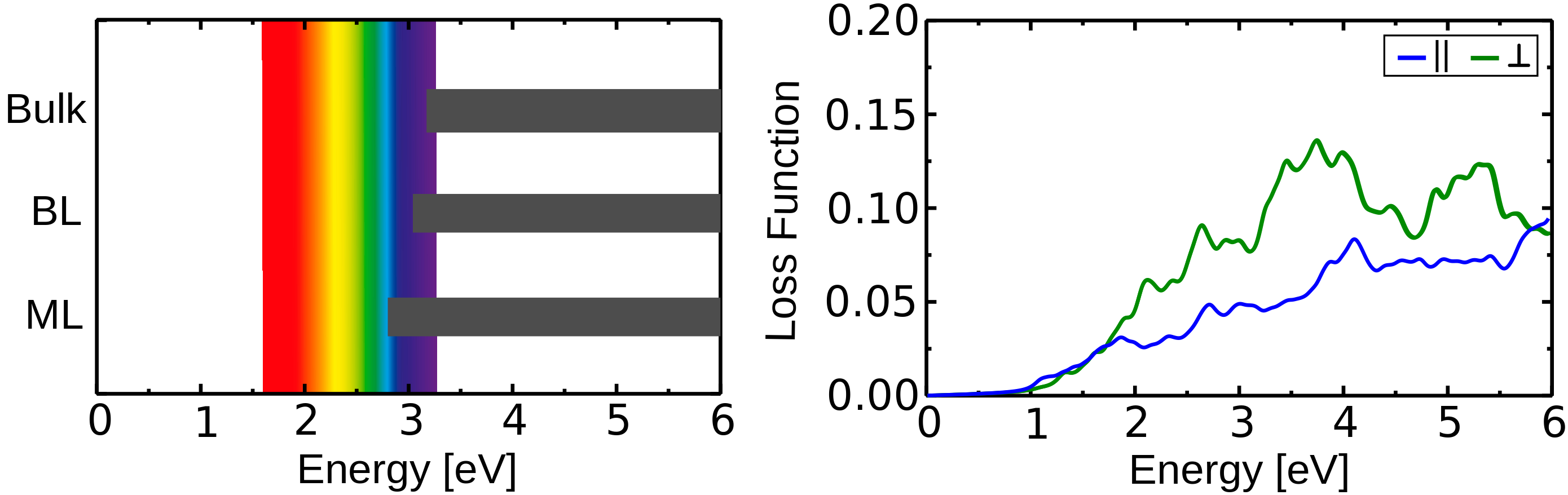}
\caption{ Hyperbolic energy window of MoTe$_{2}$ in ML, BL, and bulk form (left).
In-plane and out-of-plane component of the loss function for bulk MoTe$_{2}$ (right). A Lorentzian broadening of 82 meV accounts for excitation lifetimes.
\label{fig:Loss}}
\end{figure*}

In the bulk, the out-of-plane component exhibits  a similar absorption onset and overall analogous features compared to the in-plane one: Absorption maxima appear just below 2 eV, at about 2.5 eV, and at around 3.1 eV. In other words, bulk MoTe$_{2}$ absorbs UV light in the out-of-plane direction as much as or even slightly more than in the in-plane direction, depending on the specific wavelength. Conversely, due to the reduced dimensionality, the spectra of the BL and especially of the ML exhibit extremely weak absorption in the out-of-plane direction across the entire visible region (see also inset in Fig.~\ref{fig:Im_Re}). The pronounced anisotropy of these systems gives rise to non-negligible local-field effects (LFE) which further contribute to blue-shift the absorption, as discussed extensively in the context of TMDCs~\cite{fara+12prb,cudazzo2013local}. Most importantly, the optical selection rules inhibit light absorption in the out-of-plane direction, as recently discussed in detail by Guilhon and coworkers for a number of 2D materials including TMDCs~\cite{PhysRevB.99.161201}. Experimental evidence of these selection rules in TMDCs was provided in earlier works~\cite{zhan+15prl,funke2016imaging,wang+17prl}.

The features in the imaginary part of the macroscopic dielectric function are reflected also in the real parts, shown in the bottom left of Fig.~\ref{fig:Im_Re}. The in-plane components of Re$\epsilon_M$ change sign around 3 eV in all three phases, assuming negative values starting from 2.85 eV, 3.06 eV, and 3.23 eV, respectively, for the ML, the BL, and the bulk, and remain negative until 6 eV. The real part of the out-of-plane component of $\epsilon_M$ again reflects the features discussed for the imaginary part. In the bulk, the dielectric function remains positive in the whole spectral range from 0 to 6 eV (see Fig.~\ref{fig:Im_Re}, bottom right). In the BL and in the ML, Re$\epsilon_M$ is  positive and almost constant with values in a range of 4 to 5. As shown in the inset, some structure appears just below 5 eV, where the imaginary part has a steep increase (see Fig.~\ref{fig:Im_Re}, bottom left).

The positive values of the out-plane components together with the negative values of the in-plane counterparts makes MoTe$_{2}$ a type-II hyperbolic material in the UV-vis region, regardless of the number of layers. Figure.~\ref{fig:Loss} (left) shows the hyperbolic energy window of MoTe$_{2}$ from ML to bulk. Thereby the hyperbolic window is red-shifted by about 400 meV with increasing number of layers. 

The promising characteristics of MoTe$_{2}$ as natural type-II hyperbolic material based on the analysis of the dielectric function need to be substantiated by the analysis of the loss function. The essential prerequisite for a (natural) hyperbolic material to be used in photonic and opto-electronic applications is, as discussed above, to exhibit low energy loss at $\mathbf{q} = 0$ in the window of hyperbolicity. As an example, we show the results for the bulk material in Fig.~\ref{fig:Loss} (right). The out-of-plane component of the loss function is slightly larger than the in-plane one, however, both components show small values in the entire hyperbolic energy window. It should also be noted that in the examined energy window the material does not exhibit a proper plasmon peak, which would show up by a change of sign in the real part of the dielectric function (see also Fig.~\ref{fig:Im_Re}, bottom panels).

\begin{figure*}[th]
\includegraphics[width=0.8\textwidth]{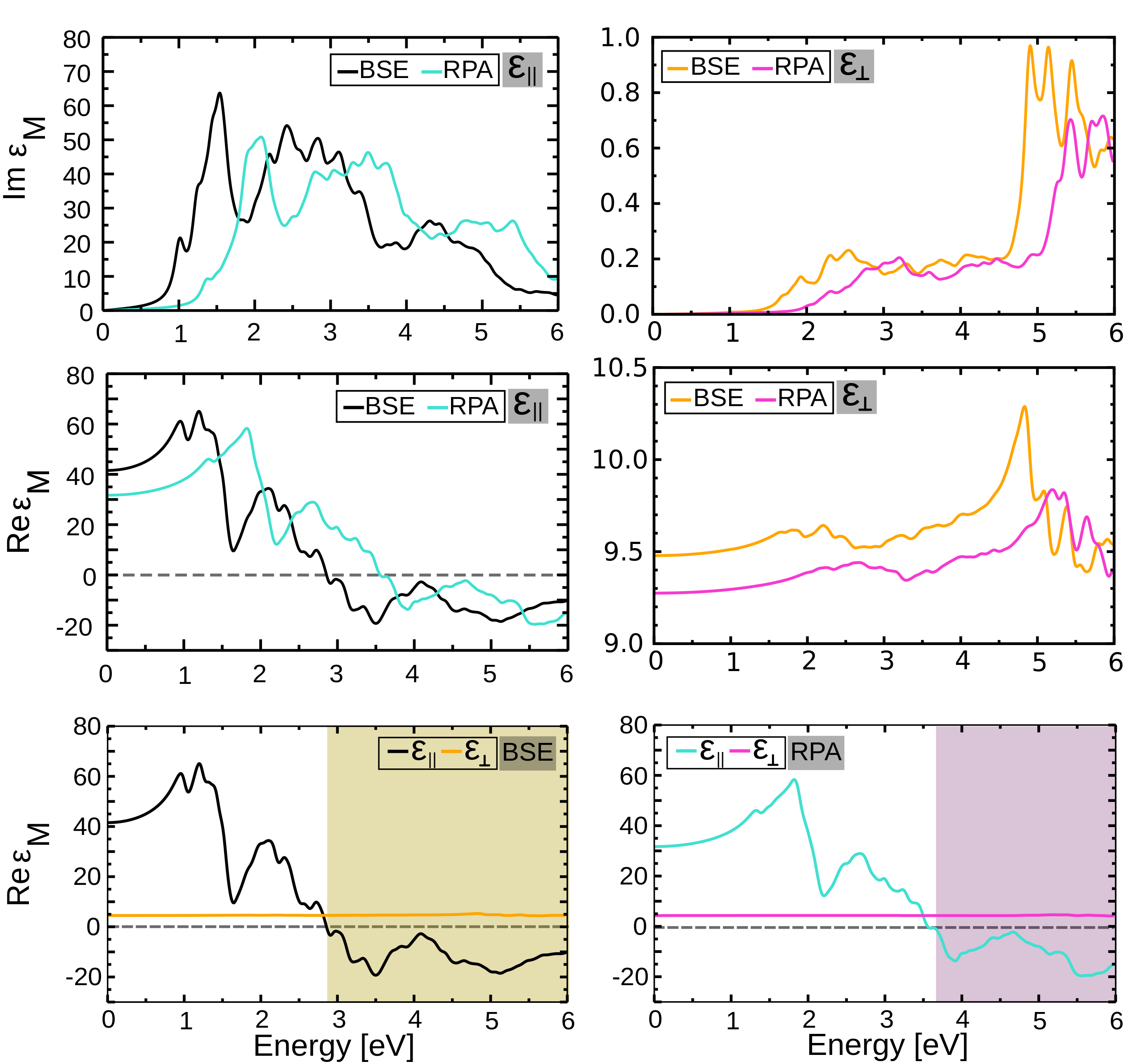}
\caption{Imaginary and real part of the macroscopic dielectric function of ML MoTe$_{2}$ obtained including excitonic effects (BSE) and neglecting them (RPA). In the bottom panels, the shaded areas highlight the hyperbolic window. A Lorentzian broadening of 82 meV is applied to account for excitation lifetimes.
\label{fig:RPA}}
\end{figure*}

The relevance of excitonic effects in the low-energy part of the absorption spectrum of TMDCs is well known \cite{berkelbach2013theory, chernikov2014exciton, li2014measurement, wang2018colloquium, druppel2018electronic}. It is thus also important to evaluate whether and to which extent these effects impact the hyperbolic dispersion of MoTe$_{2}$ over the entire UV-vis range. For this purpose, we compare in Fig.~\ref{fig:RPA} the results for the MoTe$_{2}$ ML obtained from the BSE with those of the RPA, where in both cases the QP-corrected band structure including SOC is the starting point. The imaginary part of both components is red-shifted when excitonic effects are accounted for, with the energy difference between the RPA and BSE absorption onsets being approximately 0.65 eV. This energy difference increases to 1 eV when LFE are excluded, i.e. in the independent-particle approximation (not shown). Hence, LFE play an important role but their inclusion is not sufficient for a proper description of hyperbolic behavior. The key ingredient for this purpose is an accurate treatment of the electron-hole interaction. In fact, in the real part of $\epsilon_{\parallel}$ (Fig.~\ref{fig:RPA}, middle left panel), the RPA leads to a significant overestimation (by 0.65 eV) of the energy at which Re$\epsilon_M$ crosses zero. As a consequence, RPA would (inaccurately) predict a narrower operating-frequency window in the UV, i.e. only from 3.7 to 6 eV. A previous study on the RPA level  \cite{thygesen2017calculating} reported a discontinuity of the hyperbolic window, in addition, in the MoTe$_{2}$ bulk and its heterostructure with TaS$_{2}$. However, BSE results, by including excitonic effects, predict the hyperbolic behavior of this material in the UV-vis energy window from 2.85 to 6 eV. The results shown in the bottom panels of Fig.~\ref{fig:RPA} demonstrate that the neglect of excitonic effects not only impacts the peak positions and oscillator strength in the optical absorption but can even lead to a wrong qualitative behavior for certain properties like, in this case, hyperbolicity. Its energy window in the different MoTe$_{2}$ phases, as predicted from our BSE calculations, are summarized in Fig.~\ref{fig:Loss} (left).  Compared to the bulk, the low-dimensional structures exhibit hyperbolicity in a window extended to the very upper boundary of the visible band. Overall, the hyperbolic behavior of the material regardless of its thickness is an encouraging indication towards the applicability of MoTe$_{2}$ also beyond the bulk form, as for instance in layered heterostructures.

\section{\label{sec:level1}Summary and Conclusions}

In summary, we have investigated the hyperbolic behavior of MoTe$_{2}$ as a function of dimensionality. Irrespective of the material thickness, the optical anisotropy in combination with pronounced excitonic effects leads to opposite signs of the in-plane and out-of-plane components of the dielectric function in the UV-vis region, thus generating hyperbolic dispersion. Our results are an exciting finding in view of the fact that so far, most of the natural hyperbolic materials are very sensitive to the number of layers, and previous studied reported this behavior only in the bulk \cite{esslinger2014tetradymites} or the few-layer form \cite{guo2018hyperbolic}. The periodicity in the stacking direction blue-shifts the energy window of hyperbolicity by a few tenths of an eV compared to the low-dimensional structures. The important role of excitonic effects in the hyperbolic behavior is another important result of this work. By modifying the peak positions and oscillator strength of the optical absorption they crucially impact the hyperbolic behavior, and leaving them out even leads to a qualitatively wrong behavior. 

Matching the growing interest in van der Waals heterostructures and the constantly improving techniques to synthesize materials with atomic precision, our results open new perspectives towards promoting TMDCs in the fields of photonics, opto-electronics, and nano-imaging. 
 
\begin{acknowledgments}
This work was funded by the Deutsche Forschungsgemeinschaft (DFG), projects 403180436 (ED 293/2-1) and 182087777 - SFB 951. C.C. acknowledges fruitful discussions with Arrigo Calzolari.
\end{acknowledgments}

\bibliography{MoTe2_Hyperbolic}% Produces the bibliography via BibTeX.

\end{document}